\documentclass[conference]{IEEEtran}
\IEEEoverridecommandlockouts

\usepackage[T1]{fontenc}
\usepackage{mathptmx}
\usepackage{cite}
\usepackage{amsmath,amssymb}
\usepackage{graphicx}
\usepackage{booktabs}
\usepackage{url}
\usepackage{pgfplots}
\usetikzlibrary{arrows.meta,positioning,calc}
\pgfplotsset{compat=1.18}
\usepackage{xcolor}
\definecolor{c1}{HTML}{2A78D6}
\definecolor{c2}{HTML}{EB6834}
\definecolor{c3}{HTML}{1BAF7A}
\definecolor{cBl}{HTML}{9EC5F4}
\definecolor{cB}{HTML}{1C5CAB}
\usepackage[hidelinks]{hyperref}

\begin{document}

\title{Attack Success Rate Is Not a Number:\\
On Measurement Validity in Agentic AI Security Evaluation}

\author{
\IEEEauthorblockN{Chetan Pathade}
\IEEEauthorblockA{\textit{Independent Researcher}\\
Washington, DC, USA\\
chetanpathadeur@gmail.com}
\and
\IEEEauthorblockN{Prathamesh Pawar}
\IEEEauthorblockA{\textit{Independent Researcher}\\
San Jose, CA, USA\\
prathameshpawar1301@gmail.com}
\and
\IEEEauthorblockN{Shubham Patil}
\IEEEauthorblockA{\textit{Independent Researcher}\\
San Jose, CA, USA\\
patil.pshubham@gmail.com}
}

\maketitle

\begin{abstract}
Attack success rate (ASR) is the headline metric in nearly every published
evaluation of attacks on, and defenses for, LLM agents. We argue that ASR as
currently used is not a single quantity but a family of metrics parameterized
by six design choices that papers seldom specify and never hold constant
across the literature. We support this with two studies that require no
proprietary access. First, a full-text meta-analysis of 259 agentic-security
papers posted to arXiv between February 2025 and September 2026 finds that
most report neither a variance estimate nor repeated runs for their headline
attack metric: 58\% (95\% CI 44--71) in a hand-coded random sample of 50,
65.3\% by automated coding of all 259. Only 30.9\% disclose enough about
decoding to establish whether their evaluation was even stochastic, and of
the 64 papers we confirm use an LLM judge, 29.7\% report any agreement check
against human labels. Second, an analytical study shows that these omissions are not
cosmetic: on a 100-instance benchmark, the minimum difference in ASR detectable at
conventional power is 18.2 percentage points, and two defenses whose true ASRs
differ by 5 points are ranked in the wrong order by a single-run evaluation
roughly 21\% of the time. Because several of the six axes shift ASR in a
\emph{system-dependent} way, the resulting incomparability is not a constant
offset that cancels in comparison. We conclude that cross-paper ASR comparison
is currently unsupported, and propose a ten-item reporting checklist
targeted at each failure we measure.
Our aim is not to dispute any individual result but to supply the shared
measurement contract the field has so far done without.
\end{abstract}

\begin{IEEEkeywords}
LLM agents, agentic AI, prompt injection, red teaming, evaluation methodology,
measurement validity, reproducibility
\end{IEEEkeywords}

\section{Introduction}

Research on the security of LLM agents has grown very quickly. Since it was
shown that instructions in a model's input can override its operator
\cite{perez2022ignore}, and that tool-integrated models can be compromised by
content they merely retrieve \cite{greshake2023not}, the area has produced
dedicated
benchmarks \cite{debenedetti2024agentdojo,zhan2024injecagent,zhang2024agentsecbench,andriushchenko2024agentharm},
systematizations of the agentic attack surface
\cite{sok2026,survey2026,landscape2026,pi2026coding}, ecosystem measurement
studies of emerging tool and skill distribution channels
\cite{hasan2025mcp,rethinking2026,drift2026,skillmd2026},
executable red-teaming harnesses \cite{redagentbench2026,siraj2025,seclaw2026},
and defenses ranging from runtime filters to designs with security properties
argued from construction \cite{debenedetti2025camel}.

Almost all of this work reports its central empirical claim as an
\emph{attack success rate}: the fraction of adversarial attempts that achieved
the attacker's objective, with and without some defense. The appeal is
obvious. ASR is a single bounded number, and it appears to license exactly the
comparison a reader wants to make: defense $A$ reduces ASR to 12\%, defense
$B$ only to 31\%, therefore $A$ is better.

This paper argues that the comparison is usually not licensed, because the two
numbers are not measurements of the same thing.

ASR was inherited from the single-turn jailbreak literature
\cite{wei2023jailbroken,mazeika2024harmbench}, where the unit of analysis is
unambiguous - one prompt, one response - and success is close to directly
observable. Neither property survives the move to agents. A trajectory is
multi-step, so it is unclear whether the denominator counts injections, tasks,
or trajectories. Success is mediated by side effects on an environment rather
than by the text of a reply, so it requires an oracle, and the choice of
oracle is itself a research decision that moves the number. Agent policies are
sampled stochastically, so a single run is a draw rather than a measurement.
And the attacker is no longer a fixed prompt set but a procedure whose
strength depends on its budget and on what it knows about the deployed
defense.

Each of these is a free parameter. Fixing them differently yields different
numbers from the same underlying system. When papers do not report how they
fixed them - and we find most do not - the resulting values cannot be
placed on a common axis.

This is a familiar pattern. The adversarial examples literature passed through
a period in which many published defenses were later shown to have been
evaluated under conditions that overstated their robustness
\cite{athalye2018obfuscated,tramer2020adaptive}, and the corrective was not a
new defense but a shared evaluation methodology
\cite{carlini2019evaluating}. The jailbreak literature underwent a smaller
version of the same correction when it was shown that widely used success
criteria admitted low-quality attacks \cite{souly2024strongreject}. We believe
agentic AI security is at the equivalent point.

\medskip\noindent\textbf{Contributions.}
\begin{itemize}
  \item We identify six axes along which published ASR measurements differ,
        give each an operational definition, and argue that four of them shift
        ASR in a system-dependent way, so that the induced \emph{ranking} of
        defenses is not invariant (Section~\ref{sec:axes}).
  \item We report a full-text meta-analysis of 259 agentic-security papers,
        coded by validated automatic detectors, quantifying how often each
        practice is followed (Section~\ref{sec:meta}).
  \item We give an analytical account of when measurement choices reverse a
        ranking, including minimum detectable differences for realistic
        benchmark sizes (Section~\ref{sec:reeval}).
  \item We propose a ten-item reporting checklist, each item traceable to a
        failure we measure (Section~\ref{sec:checklist}).
\end{itemize}

\medskip\noindent\textbf{What this paper does not claim.}
We do not claim that any paper in our corpus is wrong or that any defense is
ineffective. Every paper we examine is internally consistent on its own terms.
Our claim concerns the \emph{relation between} papers: the field standardized
on a metric name without standardizing on the procedure behind it, which is a
collective coordination failure rather than an individual one. We report
aggregate rates rather than naming individual papers as deficient.

\section{Background}
\label{sec:background}

A paper reporting ASR is estimating
\begin{equation}
\widehat{\mathrm{ASR}} \;=\; \frac{1}{|U|}\sum_{u \in U}
   \mathcal{O}\!\left(\tau_u\right),
\label{eq:asr}
\end{equation}
where $U$ is a set of evaluation units, $\tau_u \sim \pi_\theta(\cdot\,|\,u)$
is a trajectory sampled from the agent policy, and
$\mathcal{O}:\tau \mapsto \{0,1\}$ is a success oracle.

In the single-turn setting this is well posed: $U$ is a prompt set, $\tau$ is
one response, and $\mathcal{O}$ can be a refusal classifier whose failure
modes are well studied \cite{souly2024strongreject,mazeika2024harmbench}. In
the agentic setting all three become researcher-chosen. Benchmarks differ in
whether they enumerate injections, tasks, or sessions
\cite{debenedetti2024agentdojo,zhan2024injecagent,zhang2024agentsecbench}, and
in whether success is read off the agent's text or the environment it acted
on; and because agent evaluation involves long stochastic rollouts,
run-to-run variation is substantial even before an adversary is introduced
\cite{yao2024taubench}. Statistical practice has been examined for
language-model capabilities \cite{miller2024adding,biderman2024lessons} but
not, to our knowledge, for agentic security.

\section{Six Axes of Measurement Divergence}
\label{sec:axes}

Equation~\eqref{eq:asr} has four free parameters: the evaluation set $U$, the
oracle $\mathcal{O}$, the sampling of each $\tau_u$, and the attacker that
generates $U$. None has a canonical setting. We decompose these into six axes,
stated so that each can be coded from a paper (Section~\ref{sec:meta}) and
reasoned about analytically (Section~\ref{sec:reeval}).

The weak version of our claim is that these choices shift ASR. That alone
would be tolerable, since a uniform shift preserves the ordering of systems,
which is what readers consume. The stronger claim, which we make for the unit
of analysis (A1), the oracle (A2), attacker adaptivity (A4) and
defense-awareness (A5) but not for trial count (A3) or partial-success
binarization (A6), is that the shift is \emph{system-dependent}, so the
induced ranking is not invariant to the measurement choice.

\subsection{A1: Unit of Analysis}
Three populations are in common use: injection attempts, tasks, and
trajectories. Consider a benchmark of $T$ tasks each embedding $k$ injection
points. An attacker that reliably succeeds at exactly one injection per task
scores $1/k$ per-injection and $1.0$ per-task - a factor-$k$ gap from
identical behavior. Existing benchmarks differ on exactly this point: some
enumerate attack instances against a smaller set of underlying tasks
\cite{zhang2025msb,zhang2024agentsecbench}, others are organized around
task--injection pairs within a simulated environment
\cite{debenedetti2024agentdojo,zhan2024injecagent}.

Neither unit is wrong; they answer different questions. Per-injection ASR
estimates the reliability of an attack primitive; per-task ASR estimates the
probability that a session is compromised, usually the security-relevant
quantity since one success suffices. The two are not monotonically related
across defenses: one that blocks most injections but systematically misses a
narrow class looks strong per-injection and weak per-task.

\subsection{A2: Success Oracle}
$\mathcal{O}$ must decide whether the attacker's objective was met. Four
families appear: string or regex matching, LLM judges, environment-state
oracles, and human adjudication. They disagree in both directions. An agent
may emit the attacker's target string while never issuing the tool call that
would cause harm - a success under matching, a failure under a state oracle.
Conversely, an agent may exfiltrate data through a call whose arguments never
contain the target string.

LLM judges \cite{zheng2023judging} add a failure mode of their own:
sensitivity to the judge model, its prompt, and trajectory length. Judge-based
ASR is a measurement taken through an instrument whose calibration is seldom
reported; we find that 29.7\% of judge-using papers report agreement with
human labels (Section~\ref{sec:meta}). This matters beyond noise. Defenses
change the distribution of trajectories - more refusals, more hedging, more
truncation - so an uncalibrated judge's error rate is not constant across
the systems being compared. That is precisely the condition under which
ranking fails to be preserved, as we make quantitative in
Section~\ref{sec:oracle-inv}.

\subsection{A3: Trials and Non-determinism}
Agent policies are typically sampled at temperature $>0$, so a trajectory is a
draw, not a measurement. Under Eq.~\eqref{eq:asr} with $m=|U|$ units evaluated
once each, $\widehat{\mathrm{ASR}}$ is a binomial proportion with standard
error $\sqrt{p(1-p)/m}$, and repeated evaluation would additionally expose
run-to-run variance that a single pass cannot separate from unit-to-unit
variance. The consequences for statistical power are severe and are quantified
in Section~\ref{sec:power}. Unlike A1 and A2 this does not bias the ordering
systematically; it randomizes it, which is worse for individual small claims
and less corrosive in aggregate. Comparable concerns have been raised for
language-model capability evaluation \cite{miller2024adding,biderman2024lessons}.

\subsection{A4: Attacker Adaptivity}
ASR is a property of an (attacker, defense) pair, never of a defense alone. A
static template set lower-bounds the ASR achievable by an adaptive attacker
under budget $B$, and the \emph{looseness} of that bound is defense-dependent:
a defense keyed to surface features of known templates appears strong under
static evaluation and degrades sharply under mild rephrasing, while one with
no template sensitivity is unaffected. Static ASR therefore does not preserve
the ordering induced by adaptive ASR - the agentic restatement of a lesson
the adversarial robustness literature learned at cost
\cite{athalye2018obfuscated,tramer2020adaptive,carlini2019evaluating}. Recent
agentic work concurs, reporting that defenses effective under fixed attack
suites lose much of their advantage under adaptive pressure
\cite{abdelnabi2026always,oob2026}.

\subsection{A5: Attacker Knowledge of the Defense}
We distinguish an attacker unaware of the defense, one that knows a defense is
present, and one with access to its prompt, classifier, or policy. The middle
case is most often left unstated and is the most consequential, because a
refusal is itself an informative signal: a defense that emits a distinctive
refusal hands an adaptive attacker a cheap membership oracle over its own
decision boundary. Two defenses with identical measured ASR against an unaware
attacker can differ sharply once the attacker may iterate against that signal.
Designs that argue security from construction rather than from measured ASR
\cite{debenedetti2025camel} are partly a response to this instability.

\subsection{A6: Binarization of Partial Success}
Collapsing $\mathcal{O}$ to a bit discards structure that papers resolve
differently and rarely document. Three cases recur: partial completion, where
the agent exfiltrates some but not all of a target; refused-then-retried,
where the agent declines, the attacker escalates within the same trajectory,
and the agent complies; and recognition without execution, where the agent
articulates the risk and proceeds anyway - a pattern reported as common
enough to deserve its own name \cite{redagentbench2026}. Whether each counts
toward the numerator is a threshold choice; we find it stated in 2.3\% of
papers. Benchmarks that grade partial harm rather than binarizing
\cite{andriushchenko2024agentharm} are the exception.

\medskip\noindent\textbf{Interactions and scope.}
The axes are not independent - A2 and A6 jointly determine the numerator,
A4 and A5 jointly define the threat model - so their effects need not
compose additively, and we make no claim that they do. Nor is the list
exhaustive; model version drift, environment and tool-set versioning, and
context-window effects are further sources of incomparability we do not
measure (Section~\ref{sec:limitations}).

\section{Meta-Analysis of Reporting Practice}
\label{sec:meta}

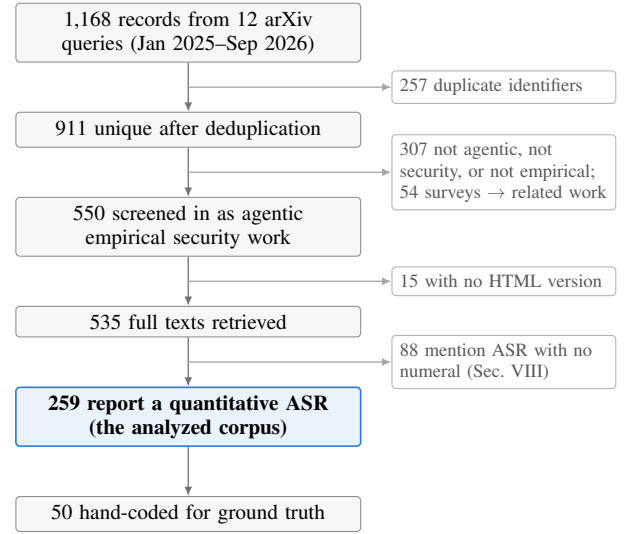
\begin{figure}[t]
\centering
\begin{tikzpicture}[
  font=\footnotesize,
  box/.style={draw=black!45, line width=0.4pt, rounded corners=1.5pt,
              align=center, inner sep=3pt, text width=4.35cm, fill=black!3},
  hit/.style={box, fill=c1!10, draw=c1, line width=0.6pt},
  side/.style={draw=black!30, line width=0.4pt, rounded corners=1.5pt,
               align=left, inner sep=3pt, text width=2.75cm, fill=none,
               font=\scriptsize, text=black!65},
  ar/.style={-{Latex[length=3.5pt,width=3pt]}, draw=black!55, line width=0.5pt},
  sar/.style={ar, draw=black!35},
]
\node[box] (n1) {1{,}168 records from 12 arXiv\\queries (Jan 2025--Sep 2026)};
\node[box, below=6.5mm of n1] (n2) {911 unique after deduplication};
\node[box, below=6.5mm of n2] (n3) {550 screened in as agentic\\empirical security work};
\node[box, below=6.5mm of n3] (n4) {535 full texts retrieved};
\node[hit,  below=6.5mm of n4] (n5) {\textbf{259 report a quantitative ASR}\\\textbf{(the analyzed corpus)}};
\node[box,  below=6.5mm of n5] (n6) {50 hand-coded for ground truth};
\foreach \a/\b in {n1/n2, n2/n3, n3/n4, n4/n5, n5/n6} \draw[ar] (\a) -- (\b);

\foreach \a/\b/\nm in {n1/n2/e1, n2/n3/e2, n3/n4/e3, n4/n5/e4}
  \coordinate (\nm) at ($(\a.south)!0.5!(\b.north)$);
\node[side, anchor=west] (s2) at ([xshift=4mm]e1 -| n1.east)
  {257 duplicate identifiers};
\node[side, anchor=west] (s3) at ([xshift=4mm]e2 -| n1.east)
  {307 not agentic, not\\security, or not empirical;\\54 surveys $\rightarrow$ related work};
\node[side, anchor=west] (s4) at ([xshift=4mm]e3 -| n1.east)
  {15 with no HTML version};
\node[side, anchor=west] (s5) at ([xshift=4mm]e4 -| n1.east)
  {88 mention ASR with no\\numeral (Sec.~\ref{sec:limitations})};
\foreach \e/\s in {e1/s2, e2/s3, e3/s4, e4/s5} \draw[sar] (\e) -- (\s.west);
\end{tikzpicture}
\caption{Corpus construction. Right-hand boxes are records removed at each
stage.}
\label{fig:corpus}
\end{figure}

\subsection{Corpus}
Fig.~\ref{fig:corpus} summarizes the construction of the corpus.
We queried the arXiv API with twelve topical queries covering prompt
injection, agentic security, tool and protocol ecosystems, and agent red
teaming, restricted to submissions between 1 January 2025 and 30 September
2026. This returned 1{,}168 records, 911 unique after deduplication by arXiv
identifier. An abstract-level screen retained papers that are both agentic
(the system takes actions: tools, protocols, skills, browsing, code
execution) and empirical security work, yielding 550 candidates and setting
aside 54 surveys and systematizations for Section~\ref{sec:related}. We
retrieved rendered full text for 535 of the 550 (97.3\%); the remaining 15
have no HTML rendering available. Of the 535, 259 met the inclusion
criterion of reporting a \emph{quantitative} attack success rate, detected as a numeric percentage within a short window of
an ASR mention. Included papers span 2025-02-27 to 2026-09-17, with 43 from
2025 and 216 from 2026 - itself an indication of how fast the area is
moving.

\subsection{Coding, and Validating the Coders}
\label{sec:validation}
Hand-coding 259 full texts was beyond our resources, so each axis is coded by
a regular-expression detector over full text. Detectors err in both
directions, so we treat their accuracy as a measurement problem of our own.
For each validated detector we drew a seeded random sample of matches,
adjudicated them by reading the surrounding context, and revised the detector
where adjudication disagreed. Table~\ref{tab:validation} records what this
changed; the constraint each revision added is stated below, so the detectors
can be reimplemented from this description.

Three failure modes \emph{inflate} a naive detector, and each is a caution
for anyone assembling such rates; deflation, the harder error to notice, is
treated in Section~\ref{sec:handcode}. \emph{Hypothetical and negated clauses} inflate every detector:
our first variance detector matched limitations sections stating that
confidence intervals \emph{would} clarify uncertainty. \emph{Off-target
metrics} inflate them further: bootstrap intervals reported for an auxiliary
classifier, not for the attack metric. Most damaging, \emph{related-work
prose} describes practices the paper does not follow - the naive adaptive
detector fired on sentences such as ``no defense is robust to adaptive
adversaries [citation]'', and on bibliography entries. We therefore require
variance and utility markers to occur near the attack metric and near a
reported numeral, and require the threat-model markers (adaptivity, budget,
defense-awareness, partial success) to appear in first-person methodological
context, away from citation markers and outside hypothetical clauses. The
last constraint alone moved four rates by 5 to 23 points.

Validation also corrected an error in the opposite direction. Our first
judge-calibration detector demanded an explicit agreement or $\kappa$
statistic and reported 13.0\%; sampling papers it coded \emph{negative}
revealed that several validate their oracle by manually inspecting its labels
without ever writing ``agreement''. Admitting that phrasing, while excluding
human-in-the-loop \emph{defenses}, which are not oracle validation, raised
the rate to 29.7\%.

Two detectors we could not rescue, and we report no rate for them. Automatic
classification of the \emph{oracle family} proved unreliable: ``environment
state'' and ``side effect'' occur constantly in agentic papers to describe
agent mechanisms and defenses rather than success criteria, and
string-matching language most often described a guardrail or prior work
rather than the study's own oracle. The difficulty is informative in itself.
Across 259 papers we could not mechanically locate a statement of how attack
success is decided in a form reliable enough to count - a comment on how
rarely that decision is stated explicitly - and we decline to convert that
into a percentage.

\begin{table}[t]
\caption{Detector Validation}
\label{tab:validation}
\centering
\footnotesize
\begin{tabular}{lccrr}
\toprule
Detector & Samp. & Prec. & Before & After \\
\midrule
Variance          & 50 & P92 R85 & 34.7 & 23.9 \\
Repeated runs     & 50 & P85 R69 & 10.8 & 20.5 \\
Judge calibration & 10p/12n & 8.5/10 & 13.0 & 29.7 \\
Uses a judge      & 10p/5n & 3/6$\to$3/4 & 41.7 & 24.7 \\
Adaptive attack   & 11p/12n & 2/6$\to$4/5 & 53.3 & 30.5 \\
Utility reported  & \phantom{0}5p & 2--3/5 & 42.5 & 37.8 \\
Temperature       & \phantom{0}5p & 5/5 & 20.8 & 20.8 \\
Seeds             & \phantom{0}4p & 4/4 & 15.1 & 15.1 \\
Code released     & \phantom{0}5p & 5/5 & 65.3 & 65.3 \\
State oracle      & \phantom{0}8p & 0/4$\to$3/4 & 43.6 & \emph{drop} \\
String matching   & \phantom{0}8p & 3/4$\to$1--2/4 & 20.5 & \emph{drop} \\
\bottomrule
\end{tabular}

\vspace{7pt}
\parbox{0.97\columnwidth}{\scriptsize \textit{Before}: the naive keyword
detector's rate. \textit{After}: the rate we report. Precision is on sampled
positives, $\to$ giving values before and after revision. The two A3
detectors were checked against the 50-paper ground-truth sample, so recall is
known for them; \emph{drop} marks a detector we could not make reliable.}
\end{table}

\subsection{Validating the Rate, Not Only the Detector}
\label{sec:handcode}
Precision on sampled positives says nothing about the papers a detector
called negative, so it cannot validate a \emph{rate}. We therefore hand-coded
a random sample of 50 included papers (fixed seed 1234) for the two
constructs behind our headline claim, reading every candidate passage and
recording an adjudication and a justification for each paper.

The exercise changed the result. Our first repeated-runs detector had perfect
precision but 44\% recall, missing phrasings such as ``each scenario was
executed three times'' and ``each method is run with three seeds'' while
correctly ignoring benchmark totals such as ``3{,}510 trials''. Broadening it
raised corpus-wide repeated-runs reporting from 10.8\% to 20.5\% and lowered
our headline from 71.0\% to 65.3\%. Against the hand-coded sample the revised
detectors give 24.0\% versus 26.0\% (variance) and 26.0\% versus 32.0\%
(repeated runs), agreeing with adjudication on 44 of 50 papers for the
composite ($\kappa=0.75$; $\kappa=0.84$ for variance and $0.66$ for repeated
runs taken separately).

We report the hand-coded figure as primary: \textbf{58.0\% (95\% Wilson CI
44.2--70.6)} of papers report neither a variance estimate nor repeated runs.
Because the detectors were revised in light of this sample, their agreement
on it is optimistic by construction; the corpus-wide 65.3\% should be read as
an automated estimate consistent with the sampled one, not as an independent
confirmation of it.

\subsection{Results}

\begin{table}[t]
\caption{Reporting Practice, 259 Papers}
\label{tab:meta}
\centering
\small
\begin{tabular}{llr}
\toprule
Axis & Practice & \% \\
\midrule
A3 & Variance or CI on attack metric (strict) & 23.9 \\
A3 & \quad same, no proximity constraint          & 34.7 \\
A3 & Repeated runs per configuration              & 20.5 \\
A3 & \textbf{Neither variance nor repeated runs}  & \textbf{65.3} \\
A3 & Any decoding info (temp./greedy/seed)   & 30.9 \\
A3 & \quad temperature stated                     & 20.8 \\
A3 & \quad seed stated                            & 15.1 \\
\midrule
A2 & Uses an LLM judge (lower bound)              & 24.7 \\
A2 & \quad of those, validated against humans     & 29.7 \\
\midrule
A4 & Adaptive attacker considered                 & 30.5 \\
A4 & Attack budget stated                         & 10.8 \\
A5 & Attacker knowledge of defense stated         & 16.2 \\
A6 & Partial success addressed                    & \phantom{0}2.3 \\
\midrule
gen. & Benign-task utility reported             & 37.8 \\
gen. & Code released                            & 65.3 \\
\bottomrule
\end{tabular}

\vspace{7pt}
\parbox{0.97\columnwidth}{\scriptsize All 259 papers report a quantitative
attack success rate. Axis labels refer to Section~\ref{sec:axes};
\textit{gen.} marks a practice not tied to one axis. Every rate comes from a
detector validated against hand adjudication (Table~\ref{tab:validation}),
with precision estimated from small samples, so rates are approximate.}
\end{table}

Table~\ref{tab:meta} reports the results. Three findings stand out.

\textbf{Uncertainty is rarely quantified.} 65.3\% of papers report neither a
variance estimate on their attack metric nor repeated runs; hand-coding a
random sample of 50 puts this at 58.0\% (95\% Wilson CI 44.2--70.6). Only
30.9\% disclose enough about decoding to establish whether the evaluation was
stochastic at all. A reader therefore usually cannot tell whether a reported
difference is larger than the noise of the procedure that produced it. Given
the power analysis in Section~\ref{sec:power}, this is the most consequential
gap we measure.

\textbf{The measuring instrument is usually uncalibrated.} At least 24.7\% of
papers use an LLM judge - a lower bound, since we count only judges we can
confirm act as the evaluation oracle - and of those, 29.7\% report any
agreement check against human labels. Judge calibration is not a formality here: because defenses
systematically change the trajectory distribution, differential judge error
across compared systems is the mechanism by which oracle choice reverses
rankings (Section~\ref{sec:oracle-inv}).

\textbf{Threat models are underspecified.} Only 30.5\% evaluate an adaptive
attacker of their own, 10.8\% state an attack budget, and 16.2\% state what
the attacker knows about the defense, so reported adaptivity frequently cannot
be reproduced or compared. Partial success is addressed by 2.3\%. All four
figures fell substantially once we required the matched text to describe the
paper's own procedure rather than related work (Section~\ref{sec:validation}),
which is itself a caution about how such rates are usually assembled. Code
release, at 65.3\%, is comparatively healthy.

\section{When Measurement Choices Reverse a Ranking}
\label{sec:reeval}

The rates above matter only if the omitted information could change a
conclusion. We show analytically that it can, under assumptions that are
favorable to current practice: we assume throughout that the evaluation is
otherwise perfect, that units are independent, and that the only defect is the
one under study. Every quantity below follows from the stated closed forms
and a fixed random seed, so it can be recomputed directly from the text; no
model calls are required.

\subsection{Sampling: Power and Ranking Inversion}
\label{sec:power}

\begin{figure}[t]
\centering
\begin{tikzpicture}
\begin{axis}[
  name=top, width=\columnwidth, height=4.4cm,
  xmode=log, log basis x=10,
  xmin=25, xmax=1000, ymin=0, ymax=47,
  xlabel={}, xticklabels={},
  ylabel={MDD (pp)}, ylabel style={font=\footnotesize, yshift=-3pt},
  ytick={0,10,20,30,40},
  ymajorgrids=true, tick label style={font=\footnotesize},
  grid=major, grid style={line width=0.3pt, draw=black!12},
  axis line style={line width=0.4pt, draw=black!55},
  tick style={line width=0.4pt, draw=black!55},
  legend style={font=\footnotesize, draw=none, fill=none,
                at={(0.98,0.97)}, anchor=north east, row sep=-2pt},
  legend cell align=left, clip marker paths=true,
]
\addplot[color=c1, line width=0.9pt, mark=*, mark size=1.3pt,
         mark options={fill=c1,draw=c1}] coordinates {
(25,36.31) (30,33.15) (40,28.71) (50,25.68) (65,22.52) (80,20.30) (100,18.16)
(125,16.24) (150,14.82) (200,12.84) (250,11.48) (300,10.48) (400,9.08)
(500,8.12) (650,7.12) (800,6.42) (1000,5.74)};
\addlegendentry{$\mathrm{ASR}=30\%$}
\addplot[color=c2, line width=0.9pt, dashed, mark=square*, mark size=1.2pt,
         mark options={solid,fill=c2,draw=c2}] coordinates {
(25,38.82) (30,35.44) (40,30.69) (50,27.45) (65,24.08) (80,21.70) (100,19.41)
(125,17.36) (150,15.85) (200,13.72) (250,12.28) (300,11.21) (400,9.70)
(500,8.68) (650,7.61) (800,6.86) (1000,6.14)};
\addlegendentry{$\mathrm{ASR}=60\%$}
\draw[line width=0.4pt, draw=black!45] (axis cs:100,0) -- (axis cs:100,42);
\node[font=\footnotesize, anchor=south west, inner sep=1.5pt]
  at (axis cs:112,27) {$18.2$ pp at $m{=}100$};
\node[font=\footnotesize\bfseries, anchor=north west] at (axis cs:26,46) {(a)};
\end{axis}

\begin{axis}[
  name=bot, at={(top.below south west)}, anchor=north west, yshift=-2pt,
  width=\columnwidth, height=4.4cm,
  xmode=log, log basis x=10,
  xmin=25, xmax=1000, ymin=0, ymax=0.56,
  xlabel={benchmark size $m$ (evaluation units)},
  xlabel style={font=\footnotesize},
  ylabel={$P$(ranking inversion)}, ylabel style={font=\footnotesize, yshift=-3pt},
  ytick={0,0.1,0.2,0.3,0.4,0.5},
  yticklabels={0,.1,.2,.3,.4,.5},
  xtick={25,50,100,200,500,1000},
  xticklabels={25,50,100,200,500,1000},
  tick label style={font=\footnotesize},
  grid=major, grid style={line width=0.3pt, draw=black!12},
  axis line style={line width=0.4pt, draw=black!55},
  tick style={line width=0.4pt, draw=black!55},
  legend style={font=\footnotesize, draw=none, fill=none,
                at={(0.98,0.97)}, anchor=north east, row sep=-2pt},
  legend cell align=left,
]
\addplot[color=c1, line width=0.9pt, mark=*, mark size=1.3pt,
         mark options={fill=c1,draw=c1}] coordinates {
(25,0.4381) (30,0.4322) (40,0.4219) (50,0.4128) (65,0.4008) (80,0.3902)
(100,0.3776) (125,0.3637) (150,0.3513) (200,0.3297) (250,0.3110) (300,0.2946)
(400,0.2665) (500,0.2429) (650,0.2134) (800,0.1890) (1000,0.1621)};
\addlegendentry{$\delta=2$ pp}
\addplot[color=c2, line width=0.9pt, dashed, mark=square*, mark size=1.2pt,
         mark options={solid,fill=c2,draw=c2}] coordinates {
(25,0.3459) (30,0.3320) (40,0.3080) (50,0.2875) (65,0.2613) (80,0.2391)
(100,0.2139) (125,0.1876) (150,0.1657) (200,0.1310) (250,0.1049) (300,0.0848)
(400,0.0564) (500,0.0381) (650,0.0216) (800,0.0124) (1000,0.0061)};
\addlegendentry{$\delta=5$ pp}
\addplot[color=c3, line width=0.9pt, densely dotted, mark=triangle*,
         mark size=1.6pt, mark options={solid,fill=c3,draw=c3}] coordinates {
(25,0.2055) (30,0.1839) (40,0.1492) (50,0.1225) (65,0.0925) (80,0.0707)
(100,0.0501) (125,0.0330) (150,0.0220) (200,0.0100) (250,0.0047) (300,0.0022)
(400,0.0005) (500,0.0001) (650,0.0000) (800,0.0000) (1000,0.0000)};
\addlegendentry{$\delta=10$ pp}
\draw[line width=0.4pt, draw=black!45] (axis cs:100,0) -- (axis cs:100,0.50);
\node[font=\footnotesize\bfseries, anchor=north west] at (axis cs:26,0.55) {(b)};
\end{axis}
\end{tikzpicture}
\caption{Single-run evaluation vs.\ benchmark size $m$. (a) Minimum
detectable difference. (b) Probability of ranking two defenses in the wrong
order, for true gap $\delta$. Vertical rule: $m{=}100$.}
\label{fig:power}
\end{figure}
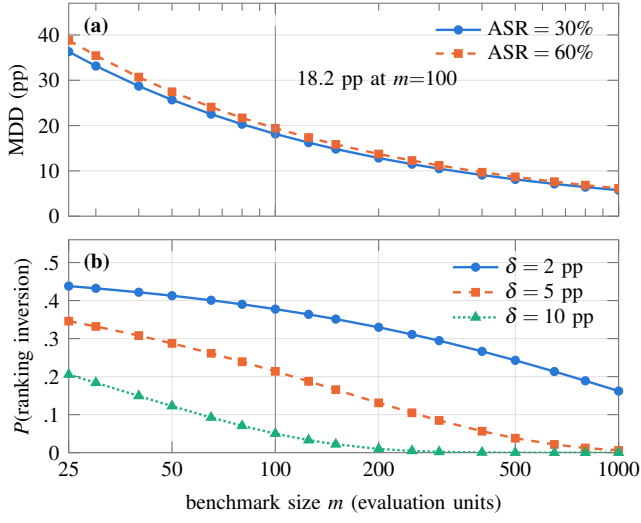

Treating $\widehat{\mathrm{ASR}}$ as a binomial proportion over $m$ units, the
minimum difference detectable at 80\% power and $\alpha=0.05$ is
$(z_{\alpha/2}+z_\beta)\sqrt{2p(1-p)/m}$. Fig.~\ref{fig:power} plots this alongside
the probability that a single-run evaluation ranks two defenses in the wrong
order.

The implications are stark. A 100-instance benchmark - a common scale in this
literature - cannot resolve differences below roughly 18 percentage points at
conventional power. Two defenses truly separated by 5 points are ranked
backwards about 21\% of the time, and by 2 points, about 38\% of the time,
which is close to a coin flip. Inversion probabilities here use a normal
approximation; a 200{,}000-trial Monte Carlo check on three cells agrees within
2.3 points and indicates the approximation slightly \emph{understates}
inversion (e.g.\ $.214$ versus $.237$ at $m{=}100$, $\delta{=}5$pp).

Combined with Table~\ref{tab:meta}, this is the paper's central practical
point: the majority of this literature reports point estimates from
single-pass evaluations at scales where differences of the size routinely
claimed are not statistically resolvable.

\subsection{Oracles: Differential Error}
\label{sec:oracle-inv}

Let an oracle have false-positive rate $\alpha$ and false-negative rate
$\beta$. The measured rate is
$\widehat{p} = p(1-\beta) + (1-p)\alpha$. For two systems with true rates
$p_A < p_B$ and oracle error $(\alpha,\beta_A)$, $(\alpha,\beta_B)$, the
ordering inverts when
\begin{equation}
\frac{1-\beta_B}{1-\beta_A} \;<\; \frac{p_A}{p_B}.
\label{eq:inv}
\end{equation}

This is a demanding condition, and we report that honestly: for $p_A=0.25$ and
$p_B=0.35$ with $\alpha=0.02$, inversion occurs in only 1 of 25 combinations
drawn from $\beta \in \{.02,.05,.10,.20,.30\}$, namely $\beta_A=0.02$ against
$\beta_B=0.30$. Oracle-induced inversion therefore requires substantial
differential error, not merely an imperfect judge.

Whether such error is plausible is the relevant question.
Equation~\eqref{eq:inv} shows the requirement scales with the relative gap:
for defenses separated by 10\% rather than 40\% in relative terms, a $\beta$
difference near 0.1 suffices. A judge's error on refusal-heavy, hedged or
truncated trajectories is exactly what differs between a defended and an
undefended system, and 70.3\% of judge-using papers report no human-agreement
check. That is the finding: not that judges do invert rankings, but that
their calibration is unmeasured in most of the literature depending on it.

\section{A Reporting Checklist}
\label{sec:checklist}

Each item is traceable to a failure measured above, and each is a sentence or
a table entry rather than an additional experiment.

\begin{enumerate}\itemsep1pt
  \item \textbf{Unit.} State whether ASR is per-injection, per-task, or
        per-trajectory, and give $|U|$. (A1)
  \item \textbf{Oracle.} State the oracle. If it is an LLM judge, give the
        judge model and prompt. (A2; we could not identify the oracle family
        reliably enough to report a rate)
  \item \textbf{Judge calibration.} Report agreement with human labels on a
        subsample. (A2; currently 29.7\%)
  \item \textbf{Trials.} Report runs per configuration; if one, say so.
        (A3; currently 20.5\% report more than one)
  \item \textbf{Variance.} Report an interval, and claim no improvement
        smaller than your MDD at the benchmark size used.
        (A3; currently 23.9\%)
  \item \textbf{Decoding.} Report temperature, top-$p$, and seeds.
        (A3; currently 30.9\% report any of these)
  \item \textbf{Attacker.} State static or adaptive, and give the query
        budget. (A4; budget currently 10.8\%)
  \item \textbf{Defense knowledge.} State what the attacker knows about the
        defense. (A5; currently 16.2\%)
  \item \textbf{Partial success.} State the binarization rule, including how
        refused-then-retried trajectories score. (A6; currently 2.3\%)
  \item \textbf{Utility.} Report benign-task success under the same defense.
        (currently 37.8\%)
\end{enumerate}

Item 10 deserves emphasis. A defense can drive ASR toward zero by refusing
broadly, and without a paired utility number a reader cannot distinguish that
from a real defense. At 37.8\%, this is the single cheapest improvement
available to the field.

\section{Related Work}
\label{sec:related}

\textbf{Evaluation methodology.} Our framing follows the adversarial
robustness correction, in which broken evaluations
\cite{athalye2018obfuscated}, adaptive re-evaluation
\cite{tramer2020adaptive} and a methodology checklist
\cite{carlini2019evaluating} reset the field's standards; the jailbreak
analog showed permissive success criteria inflate apparent attack strength
\cite{souly2024strongreject}. Statistical rigor has been treated for
capabilities \cite{miller2024adding,biderman2024lessons} and reproducibility
broadly \cite{pineau2020reproducibility}. We extend these to agentic
security, where the oracle and unit-of-analysis problems are distinctive.

\textbf{Agentic security benchmarks.} AgentDojo
\cite{debenedetti2024agentdojo}, InjecAgent \cite{zhan2024injecagent}, ASB
\cite{zhang2024agentsecbench} and AgentHarm
\cite{andriushchenko2024agentharm} established the templates the area now
uses; protocol- and ecosystem-specific benchmarks
\cite{yang2025mcpsecbench,zhang2025msb,rethinking2026} and red-teaming
harnesses \cite{redagentbench2026,siraj2025,seclaw2026} extend them. Our
corpus is largely composed of these papers, and the checklist is meant for
their successors to adopt, not to supersede them.

\textbf{Systematizations.} Recent SoKs and surveys
\cite{sok2026,survey2026,landscape2026,pi2026coding} observe qualitatively
that evaluation practice is heterogeneous, and one meta-analyzes attack
effectiveness across studies \cite{pi2026coding}. We take that observation as
our starting point and convert it into a measured claim with a defined
corpus, validated coders and an analysis of consequences. Ecosystem
measurement studies \cite{hasan2025mcp,drift2026,skillmd2026} are adjacent,
measuring artifacts rather than papers.

\section{Limitations}
\label{sec:limitations}

Our corpus is arXiv-only and English-only, and query-based retrieval will have
missed relevant work; 2.7\% of screened candidates had no HTML rendering and
are absent. Coding is automatic. We validated eleven detectors by hand
(Table~\ref{tab:validation}): six required revision, two were abandoned, and
three passed unchanged. The two detectors behind our headline claim were
additionally checked against a hand-coded random sample of the full corpus
(Section~\ref{sec:handcode}), which is why we can state recall for them and
report a confidence interval. The other rates rest on precision estimated
from small samples of matches (4--12 items) with recall unmeasured; because
low recall is the error mode we found in the one place we could measure it,
those rates are more likely understated than overstated, and we recommend
reading them as lower bounds. All adjudication is single-coder. The $\kappa=0.75$ we
report measures a detector against one adjudicator, not two humans against
each other, so it bounds automation error rather than adjudicator
subjectivity; a second independent human rater on the same 50 papers is the
clearest further improvement.

Inclusion requires a numeric ASR near an ASR mention, and we measured the cost
of that rule: 88 further papers mention an attack success rate without a
detected numeral, 26 of them exhibiting a signature of HTML-to-text numeral
loss (e.g.\ ``reduces attack success rate by over \%''). The true included
population is therefore between 259 and 347. Numeral loss is a rendering
artifact, so we expect it to be roughly independent of whether a paper reports
variance or calibrates its judge, leaving the rates in Table~\ref{tab:meta}
less affected than the count - but we have not verified that independence,
and a systematic relationship between document class and reporting practice
would bias our rates in an unknown direction.

The Section~\ref{sec:reeval} results are analytical: they show what
\emph{must} follow from binomial sampling and oracle error, under
independence assumptions real benchmarks violate - correlated tasks and
shared environments make effective sample size smaller than $m$, so our MDDs
are optimistic. We have not run defenses end to end, so we report no empirical
ranking inversion; that experiment is the natural next step and needs model
access we did not have.

Finally, the checklist is proposed, not validated. We have not shown that
adopting it improves cross-paper comparability, only that each item
corresponds to a gap we measured.

\section{Conclusion}

Agentic AI security has produced a large and fast-growing body of empirical
work built on a metric whose definition the field never standardized. Across
259 papers, most report neither variance nor repeated runs - 58\% in a
hand-coded sample, 65.3\% by automated coding - and 70.3\% of those
confirmed to use an LLM judge report no calibration of it. At the benchmark sizes
in common use, differences of the magnitude routinely claimed are not
statistically resolvable, and four of the six measurement axes we identify can
reverse a ranking rather than merely shift it. The field does not need a
better number. It needs to agree on what the number is.

\section*{Reproducibility}
The corpus is defined by twelve arXiv API queries over a stated date window
and reconstructible from them; the screening rule, the detector constraints,
and the hand-coding sample seed are all specified in the text. The analytical
results in Section~\ref{sec:reeval} follow from the closed forms given there.
We have not published an artifact repository.

\bibliographystyle{IEEEtran}
\bibliography{refs}

@article{athalye2018obfuscated,
  author  = {Anish Athalye and Nicholas Carlini and David Wagner},
  title   = {{Obfuscated Gradients Give a False Sense of Security: Circumventing Defenses to Adversarial Examples}},
  journal = {arXiv preprint arXiv:1802.00420},
  year    = {2018},
  url     = {https://arxiv.org/abs/1802.00420}
}

@article{carlini2019evaluating,
  author  = {Nicholas Carlini and Anish Athalye and Nicolas Papernot and Wieland Brendel and Jonas Rauber and Dimitris Tsipras and Ian Goodfellow and Aleksander Madry and others},
  title   = {{On Evaluating Adversarial Robustness}},
  journal = {arXiv preprint arXiv:1902.06705},
  year    = {2019},
  url     = {https://arxiv.org/abs/1902.06705}
}

@article{tramer2020adaptive,
  author  = {Florian Tramer and Nicholas Carlini and Wieland Brendel and Aleksander Madry},
  title   = {{On Adaptive Attacks to Adversarial Example Defenses}},
  journal = {arXiv preprint arXiv:2002.08347},
  year    = {2020},
  url     = {https://arxiv.org/abs/2002.08347}
}

@article{pineau2020reproducibility,
  author  = {Joelle Pineau and Philippe Vincent-Lamarre and Koustuv Sinha and Vincent Larivière and Alina Beygelzimer and Florence d'Alché-Buc and Emily Fox and Hugo Larochelle},
  title   = {{Improving Reproducibility in Machine Learning Research (A Report from the NeurIPS 2019 Reproducibility Program)}},
  journal = {arXiv preprint arXiv:2003.12206},
  year    = {2020},
  url     = {https://arxiv.org/abs/2003.12206}
}

@article{perez2022ignore,
  author  = {Fábio Perez and Ian Ribeiro},
  title   = {{Ignore Previous Prompt: Attack Techniques For Language Models}},
  journal = {arXiv preprint arXiv:2211.09527},
  year    = {2022},
  url     = {https://arxiv.org/abs/2211.09527}
}

@article{greshake2023not,
  author  = {Kai Greshake and Sahar Abdelnabi and Shailesh Mishra and Christoph Endres and Thorsten Holz and Mario Fritz},
  title   = {{Not what you've signed up for: Compromising Real-World LLM-Integrated Applications with Indirect Prompt Injection}},
  journal = {arXiv preprint arXiv:2302.12173},
  year    = {2023},
  url     = {https://arxiv.org/abs/2302.12173}
}

@article{zheng2023judging,
  author  = {Lianmin Zheng and Wei-Lin Chiang and Ying Sheng and Siyuan Zhuang and Zhanghao Wu and Yonghao Zhuang and Zi Lin and Zhuohan Li and others},
  title   = {{Judging LLM-as-a-Judge with MT-Bench and Chatbot Arena}},
  journal = {arXiv preprint arXiv:2306.05685},
  year    = {2023},
  url     = {https://arxiv.org/abs/2306.05685}
}

@article{wei2023jailbroken,
  author  = {Alexander Wei and Nika Haghtalab and Jacob Steinhardt},
  title   = {{Jailbroken: How Does LLM Safety Training Fail?}},
  journal = {arXiv preprint arXiv:2307.02483},
  year    = {2023},
  url     = {https://arxiv.org/abs/2307.02483}
}

@article{mazeika2024harmbench,
  author  = {Mantas Mazeika and Long Phan and Xuwang Yin and Andy Zou and Zifan Wang and Norman Mu and Elham Sakhaee and Nathaniel Li and others},
  title   = {{HarmBench: A Standardized Evaluation Framework for Automated Red Teaming and Robust Refusal}},
  journal = {arXiv preprint arXiv:2402.04249},
  year    = {2024},
  url     = {https://arxiv.org/abs/2402.04249}
}

@article{souly2024strongreject,
  author  = {Alexandra Souly and Qingyuan Lu and Dillon Bowen and Tu Trinh and Elvis Hsieh and Sana Pandey and Pieter Abbeel and Justin Svegliato and others},
  title   = {{A StrongREJECT for Empty Jailbreaks}},
  journal = {arXiv preprint arXiv:2402.10260},
  year    = {2024},
  url     = {https://arxiv.org/abs/2402.10260}
}

@article{zhan2024injecagent,
  author  = {Qiusi Zhan and Zhixiang Liang and Zifan Ying and Daniel Kang},
  title   = {{InjecAgent: Benchmarking Indirect Prompt Injections in Tool-Integrated Large Language Model Agents}},
  journal = {arXiv preprint arXiv:2403.02691},
  year    = {2024},
  url     = {https://arxiv.org/abs/2403.02691}
}

@article{biderman2024lessons,
  author  = {Stella Biderman and Hailey Schoelkopf and Lintang Sutawika and Leo Gao and Jonathan Tow and Baber Abbasi and Alham Fikri Aji and Pawan Sasanka Ammanamanchi and others},
  title   = {{Lessons from the Trenches on Reproducible Evaluation of Language Models}},
  journal = {arXiv preprint arXiv:2405.14782},
  year    = {2024},
  url     = {https://arxiv.org/abs/2405.14782}
}

@article{yao2024taubench,
  author  = {Shunyu Yao and Noah Shinn and Pedram Razavi and Karthik Narasimhan},
  title   = {{$\tau$-bench: A Benchmark for Tool-Agent-User Interaction in Real-World Domains}},
  journal = {arXiv preprint arXiv:2406.12045},
  year    = {2024},
  url     = {https://arxiv.org/abs/2406.12045}
}

@article{debenedetti2024agentdojo,
  author  = {Edoardo Debenedetti and Jie Zhang and Mislav Balunović and Luca Beurer-Kellner and Marc Fischer and Florian Tramèr},
  title   = {{AgentDojo: A Dynamic Environment to Evaluate Prompt Injection Attacks and Defenses for LLM Agents}},
  journal = {arXiv preprint arXiv:2406.13352},
  year    = {2024},
  url     = {https://arxiv.org/abs/2406.13352}
}

@article{zhang2024agentsecbench,
  author  = {Hanrong Zhang and Jingyuan Huang and Kai Mei and Yifei Yao and Zhenting Wang and Chenlu Zhan and Hongwei Wang and Yongfeng Zhang},
  title   = {{Agent Security Bench (ASB): Formalizing and Benchmarking Attacks and Defenses in LLM-based Agents}},
  journal = {arXiv preprint arXiv:2410.02644},
  year    = {2024},
  url     = {https://arxiv.org/abs/2410.02644}
}

@article{andriushchenko2024agentharm,
  author  = {Maksym Andriushchenko and Alexandra Souly and Mateusz Dziemian and Derek Duenas and Maxwell Lin and Justin Wang and Dan Hendrycks and Andy Zou and others},
  title   = {{AgentHarm: A Benchmark for Measuring Harmfulness of LLM Agents}},
  journal = {arXiv preprint arXiv:2410.09024},
  year    = {2024},
  url     = {https://arxiv.org/abs/2410.09024}
}

@article{miller2024adding,
  author  = {Evan Miller},
  title   = {{Adding Error Bars to Evals: A Statistical Approach to Language Model Evaluations}},
  journal = {arXiv preprint arXiv:2411.00640},
  year    = {2024},
  url     = {https://arxiv.org/abs/2411.00640}
}

@article{debenedetti2025camel,
  author  = {Edoardo Debenedetti and Ilia Shumailov and Tianqi Fan and Jamie Hayes and Nicholas Carlini and Daniel Fabian and Christoph Kern and Chongyang Shi and others},
  title   = {{Defeating Prompt Injections by Design}},
  journal = {arXiv preprint arXiv:2503.18813},
  year    = {2025},
  url     = {https://arxiv.org/abs/2503.18813}
}

@article{hasan2025mcp,
  author  = {Mohammed Mehedi Hasan and Hao Li and Emad Fallahzadeh and Gopi Krishnan Rajbahadur and Bram Adams and Ahmed E. Hassan},
  title   = {{Model Context Protocol (MCP) at First Glance: Studying the Security and Maintainability of MCP Servers}},
  journal = {arXiv preprint arXiv:2506.13538},
  year    = {2025},
  url     = {https://arxiv.org/abs/2506.13538}
}

@article{yang2025mcpsecbench,
  author  = {Yixuan Yang and Cuifeng Gao and Daoyuan Wu and Yufan Chen and Yingjiu Li and Shuai Wang},
  title   = {{MCPSecBench: A Systematic Security Benchmark and Playground for Testing Model Context Protocols}},
  journal = {arXiv preprint arXiv:2508.13220},
  year    = {2025},
  url     = {https://arxiv.org/abs/2508.13220}
}

@article{zhang2025msb,
  author  = {Dongsen Zhang and Zekun Li and Xu Luo and Xuannan Liu and Peipei Li and Wenjun Xu},
  title   = {{MCP Security Bench (MSB): Benchmarking Attacks Against Model Context Protocol in LLM Agents}},
  journal = {arXiv preprint arXiv:2510.15994},
  year    = {2025},
  url     = {https://arxiv.org/abs/2510.15994}
}

@article{siraj2025,
  author  = {Kaiwen Zhou and Ahmed Elgohary and A S M Iftekhar and Amin Saied},
  title   = {{SIRAJ: Diverse and Efficient Red-Teaming for LLM Agents via Distilled Structured Reasoning}},
  journal = {arXiv preprint arXiv:2510.26037},
  year    = {2025},
  url     = {https://arxiv.org/abs/2510.26037}
}

@article{pi2026coding,
  author  = {Narek Maloyan and Dmitry Namiot},
  title   = {{Prompt Injection Attacks on Agentic Coding Assistants: A Systematic Analysis of Vulnerabilities in Skills, Tools, and Protocol Ecosystems}},
  journal = {arXiv preprint arXiv:2601.17548},
  year    = {2026},
  url     = {https://arxiv.org/abs/2601.17548}
}

@article{landscape2026,
  author  = {Peiran Wang and Xinfeng Li and Chong Xiang and Jinghuai Zhang and Ying Li and Lixia Zhang and Xiaofeng Wang and Yuan Tian},
  title   = {{The Landscape of Prompt Injection Threats in LLM Agents: From Taxonomy to Analysis}},
  journal = {arXiv preprint arXiv:2602.10453},
  year    = {2026},
  url     = {https://arxiv.org/abs/2602.10453}
}

@article{survey2026,
  author  = {Juhee Kim and Xiaoyuan Liu and Zhun Wang and Shi Qiu and Bo Li and Wenbo Guo and Dawn Song},
  title   = {{The Attack and Defense Landscape of Agentic AI: A Comprehensive Survey}},
  journal = {arXiv preprint arXiv:2603.11088},
  year    = {2026},
  url     = {https://arxiv.org/abs/2603.11088}
}

@article{sok2026,
  author  = {Ali Dehghantanha and Sajad Homayoun},
  title   = {{SoK: The Attack Surface of Agentic AI - Tools and Autonomy}},
  journal = {arXiv preprint arXiv:2603.22928},
  year    = {2026},
  url     = {https://arxiv.org/abs/2603.22928}
}

@article{skillmd2026,
  author  = {Shoumik Saha and Kazem Faghih and Soheil Feizi},
  title   = {{Under the Hood of SKILL.md: Semantic Supply-chain Attacks on AI Agent Skill Registry}},
  journal = {arXiv preprint arXiv:2605.11418},
  year    = {2026},
  url     = {https://arxiv.org/abs/2605.11418}
}

@article{abdelnabi2026always,
  author  = {Sahar Abdelnabi and Eugene Bagdasarian},
  title   = {{AI Agents May Always Fall for Prompt Injections}},
  journal = {arXiv preprint arXiv:2605.17634},
  year    = {2026},
  url     = {https://arxiv.org/abs/2605.17634}
}

@article{seclaw2026,
  author  = {Hao Cheng and Changtao Miao and Tianle Song and Yin Wu and He Liu and Erjia Xiao and Junchi Chen and Xiaoyu Shi and others},
  title   = {{SeClaw: Spec-Driven Security Task Synthesis for Evaluating Autonomous Agents}},
  journal = {arXiv preprint arXiv:2606.02302},
  year    = {2026},
  url     = {https://arxiv.org/abs/2606.02302}
}

@article{oob2026,
  author  = {Praneeth Narisetty and Shiva Nagendra Babu Kore and Uday Kumar Reddy Kattamanchi and Jayaram Kumarapu},
  title   = {{Adaptive Evaluation of Out-of-Band Defenses Against Prompt Injection in LLM Agents}},
  journal = {arXiv preprint arXiv:2606.26479},
  year    = {2026},
  url     = {https://arxiv.org/abs/2606.26479}
}

@article{rethinking2026,
  author  = {Pei Chen and Baichao An and Mengying Wu and Binwang Wan and Geng Hong and Jinsong Chen and Xudong Pan and Jiarun Dai and others},
  title   = {{Rethinking MCP Security: A Large-Scale Study of Runtime MCP Servers and Security Scanner Reliability}},
  journal = {arXiv preprint arXiv:2607.11086},
  year    = {2026},
  url     = {https://arxiv.org/abs/2607.11086}
}

@article{redagentbench2026,
  author  = {Zixing Chen and Xingyuan Liu and Jie Zhu and Huaixia Dou and Shuo Jiang and Junhui Li and Lifan Guo and Feng Chen and others},
  title   = {{REDAgentBench: Executable Red Teaming and Faithful Measurement of LLM Agent Systems}},
  journal = {arXiv preprint arXiv:2608.10669},
  year    = {2026},
  url     = {https://arxiv.org/abs/2608.10669}
}

@article{drift2026,
  author  = {Obada Kraishan},
  title   = {{Same Name, Different Server: A Security Census of Silent Drift in the Model Context Protocol Ecosystem}},
  journal = {arXiv preprint arXiv:2609.14119},
  year    = {2026},
  url     = {https://arxiv.org/abs/2609.14119}
}

\end{document}